\documentclass[ams, nofootinbib,superscriptaddress,twocolumn,showkeys
]{revtex4}
\usepackage{amstext,amsmath,amssymb,amsfonts,bbm}
\usepackage[latin1]{inputenc}
\usepackage{fancyhdr}
\usepackage{graphicx}
\usepackage{hyperref}
\usepackage{epstopdf}
\usepackage{amsthm}
\usepackage{pstricks}

\usepackage{color}

\def\be{\begin{equation}}
\def\ee{\end{equation}}
\def\bes{\begin{eqnarray}}
\def\ees{\end{eqnarray}}

\def\pp{\partial}

\def\la{\langle} \def\ra{\rangle}
\def\f{\frac}

\def\alg{\mathfrak{g}}
\def\F{\mathcal{F}}

\DeclareMathOperator{\im}{im}

\DeclareMathOperator{\Hom}{Hom}
\DeclareMathOperator{\rk}{rk}
\DeclareMathOperator{\id}{id}
\DeclareMathOperator{\SU}{SU(2)}

\def\u{\underset}

\theoremstyle{definition}
\theoremstyle{definition}
\theoremstyle{definition}
\theoremstyle{definition}
\theoremstyle{definition}
\theoremstyle{definition}

\begin{document}

\title{\large \bf Bubble divergences from cellular cohomology}

\author{Valentin Bonzom}\author{Matteo Smerlak}

\affiliation{Centre de Physique Th\'eorique, Campus de Luminy, Case 907\\ 13288 Marseille Cedex 09 France}

\date{\small\today}

\begin{abstract}\noindent
We consider a class of lattice topological field theories, among which are the weak-coupling limit of 2d Yang-Mills theory, the Ponzano-Regge model of 3d quantum gravity and discrete BF theory, whose dynamical variables are flat discrete connections with compact structure group on a cell 2-complex. In these models, it is known that the path integral measure is ill-defined in general, because of a phenomenon called `bubble divergences'. A common expectation is that the degree of these divergences is given by the number of `bubbles' of the 2-complex. In this note, we show that this expectation, although not realistic in general, is met in some special cases: when the 2-complex is simply connected, or when the structure group is Abelian -- in both cases, the divergence degree is given by the second Betti number of the 2-complex.
\end{abstract}

\maketitle


\section{Introduction}
One road to the quantization of background-independent field theories, such as Schwarz-type topological field theories or general relativity, is the spinfoam formalism (although outdated, \cite{Baez:1999sr,perez} remain good reviews; see also \cite{fk action principle}, and \cite{epr,fk} for more recent developments). In this approach, the Feynman path integral is realized as a sum of amplitudes associated to oriented two-dimensional cell complexes, aka \emph{foams}.

Spinfoam amplitudes, however, are plagued by \emph{bubble divergences}, which arise notably in the topological models whose dynamical variables are {\it flat $G$-connections}, with $G$ a compact Lie group. This includes the weak-coupling limit of 2d Yang-Mills theory, the Ponzano-Regge model of 3d quantum gravity and discrete BF theory in higher dimensions. In these instances, the foam $\Gamma$ is the $2$-skeleton of the cell complex dual to the triangulated spacetime manifold.\footnote{In these cases, the dimension of the spacetime manifold is equal to the number of faces ($2$-cells) adjacent to each edge ($1$-cell) of $\Gamma$.} Here, however, we will not restrict ourselves to these cases, and consider arbitrary foams. This is relevant both from the perspective of canonical loop quantum gravity, where foams are interpreted as `gauge histories' of spin-networks \cite{perez}, and of {\it group field theory} \cite{GFT1,GFT2}, where foams are generated as Feynman diagrams of a certain auxialiary field theory: in both cases, the one has to deal with foams which are {\it not} dual to triangulated manifolds of the relevant dimension.


Although various regularization schemes have been proposed,\footnote{For example, the Turaev-Viro and Crane-Yetter models, in three and four dimensions respectively, make use of a quantum group in the place of $G$.} the structure of these divergences has not been elucidated so far. In particular, no general result is known concerning the {\it divergence degree} of a foam $\Gamma$: the number $\Omega(\Gamma,G)$ such that
\be
0<\u{\Lambda\rightarrow 0}{\lim} \vert\Lambda^{-\Omega(\Gamma,G)}\mathcal{Z}_{\Lambda}(\Gamma,G)\vert<\infty
\ee
where $\Lambda$ is a suitable cutoff, and $\mathcal{Z}_{\Lambda}$ the corresponding regularized amplitude. Upon inspection, it appears heuristically that $\Omega(\Gamma,G)$ is related to the number of `bubbles' of $\Gamma$ -- independent sets of faces forming closed surfaces \cite{rovelli perez}. A partial result along these lines, in $3$ dimensions with structure group $\SU$, has indeed been obtained recently in \cite{fgo}, where bubbles are characterized in a purely graphical way. Specifically, they show that for a special class of foams coined `type 1', the divergence degree is given by $B-1$, where $B$ is the number of these `graphical bubbles'. Similar results are also given in \cite{ben geloun abelien}, where for the sake of simplicity the structure group $G$ is replaced by the non-compact Abelian group $\mathbb{R}$.

Our purpose in this note is to clarify, in the topological spinfoams models based on flat connections, the relation between the divergence degree of a closed foam and the number of its `bubbles'. For that matter, we introduce in section \ref{bubble} the (co)homological language suitable to make precise the notion of `bubbles'. In section \ref{divergence}, we discuss the relation between the bubbles and the divergence degree of a foam $\Gamma$, showing in particular the following results:
\begin{itemize}
\item If $\Gamma$ is simply connected or if $G$ is Abelian, then $\Omega(\Gamma,G) = \bigl(\dim G\bigr)b_2(\Gamma)$.
\item If $\Gamma$ admits a single flat connection, then $(b_2(\Gamma)+\chi(\Gamma)-1)/2\leq \Omega(\Gamma,G)/\dim G \leq  b_2(\Gamma)$.
\end{itemize}
where $b_2(\Gamma)$ and $\chi(\Gamma)$ are the second Betti number and the Euler characteristic of $\Gamma$ respectively. In the conclusion, we explain why such cellular invariants, however, are too rough to capture the divergence degree of a generic foam, and outline a finer analysis, in terms of twisted cohomology, to be detailed in a forthcoming paper \cite{twisted}.

\section{Preliminaries}\label{bubble}

Our setting is the following. Let $G$ be a compact semi-simple Lie group with Lie algebra $\alg$, and $\Gamma$ be a closed foam (i.e. an oriented cell 2-complex without boundary). We denote $\Gamma_i$ ($i=0,1,2$) the set of its $i$-cells (vertices, edges and faces respectively), and $V=\vert\Gamma_0\vert$, $E=\vert\Gamma_1\vert$, $F=\vert\Gamma_2\vert$. 

A connection on $\Gamma$ is the assignment of elements of the structure group $G$ to each edge of $\Gamma$ (`parallel transport' operators). The space of connections on $\Gamma$ is therefore
\be
\mathcal{A} = \bigl\{ A = (g_e)_{e\in\Gamma_1}\,\in G^E\bigr\},
\ee
The curvature of a connection $A$ is the family of $F$ group elements given by
\be \label{curvature map}
H(A)= \Bigl( H_f(A)=\prod_{e\in\partial f} g_e^{[f:e]} \Bigr)_{f\in\Gamma_2},
\ee
where $[f:e]$ is the incidence number of the face $f$ on the edge $e$ ($H_f(A)$ is the `holonomy' of the connection $A$ around the face $f$). A connection $A$ is flat if\footnote{In this note, ${\bf 1}$ denotes the unit element of the relevant group.}
\be
H(A)={\bf 1}.
\ee

The spinfoam amplitude considered in this note is then defined formally as the partition function of a system of flat $G$-connections on $\Gamma$:
\be \label{amplitude}
\mathcal{Z}(\Gamma,G) = \int_{\mathcal{A}} dA\prod_{f\in\Gamma_2} \delta\bigl(H_f(A)\bigr),
\ee
where $dA = \prod_{e\in\Gamma_1} dg_e$ is the Haar measure on $\mathcal{A}=G^E$, and $\delta(g)$ is the Dirac delta on $G$. Obviously, the support of this integral is the set of flat connections
\be
\F = H^{-1}({\bf 1}).
\ee

\subsection{Gauge transformations}

Gauge transformations change the local frame at each vertex of $\Gamma$: this defines an action of $G^V$ on the set of discrete connections, according to
\be
h \cdot A = \bigl(h_{t(e)}\,g_e\,h^{-1}_{s(e)}\bigr)_{e\in\Gamma_1}.
\ee
Here $h=(h_v)_{v\in\Gamma_0}$ is a set of $V$ group elements, and $t(e)$ (respectively $s(e)$) is the end (respectively starting) vertex of the edge $e$.

Gauge transformations leave the integrand of (\ref{amplitude}) invariant. When there is more than a single vertex in $\Gamma_0$, it is therefore convenient to partially fix this gauge symmetry by setting $g_e={\bf 1}$ on every edge of a maximal tree in $\Gamma$, i.e. a subgraph of $\Gamma_1$ touching every vertex of $\Gamma$ without forming any loop. This is actually equivalent to considering the partition function on a deformation retract of $\Gamma$ with only one vertex. The remaining gauge transformations consist in the standard action of $G$ on $\mathcal{A}$ by a global conjugation. Thanks to the homotopy invariance of the divergence degree, see below, we will assume without loss of generality that $\Gamma$ is indeed of this kind.

The fundamental group $\pi_1(\Gamma)$ of such a foam admits a presentation with one generator per edge, together with one relation per face exactly of the form of $H_f$, \eqref{curvature map}. It follows that the space of flat connections $\F$ can be identified with the representation variety of $\pi_1(\Gamma)$ into $G$:
\be
\F = \Hom \bigl(\pi_1(\Gamma),G\bigr).
\ee
In particular, when $\Gamma$ is simply connected, $\F$ is just a point.


\subsection{(Co)homological bubbles}

Naively speaking, the divergence of \eqref{amplitude} comes from the fact that some of the delta functions attached to the faces of the foam $\Gamma$ are redundant. This phenomenon was first observed by Ponzano and Regge \cite{PR} in the context of 3d quantum gravity, where $\Gamma$ is the $2$-skeleton of the dual cell complex to a triangulated closed 3-manifold $\Delta$ and the gauge group is $\textrm{SU}(2)$. Using a sharp cutoff $\Lambda$ on the spins labelling the irreducible representations of $\textrm{SU}(2)$ in the Peter-Weyl decomposition of the delta function, they interpreted the divergence as coming from the $N$ {\it vertices} of $\Delta$, and conjectured that $\Omega(\Gamma,\textrm{SU}(2))=3N$. This point of view was then strengthened by Freidel and Louapre \cite{freidel louapre diffeo}, who gave a geometric interpretation to this conjecture by exhibiting a discrete Bianchi identity associated to the vertices of $\Delta$: if $F_v$ denotes the faces of $\Gamma$ `wrapping around' a vertex $v$ of the triangulation $\Delta$, then there is an ordering of $F_v$ such that 
\be \label{bubble group variables}
\prod_{f\in F_v} H_f^{\epsilon_f} = {\bf 1},
\ee
where $\epsilon_f$ is $\pm1$. 

Unfortunately, this intuition -- divergences associated to vertices in three dimensions, or $(n-3)$-simplices in $n$ dimensions -- turns out to be misleading. First, as mentioned in the introduction, one would like to consider foams that are not dual to any simplicial complex, hence where the notion of $(n-3)$-simplex of the dual triangulation is meaningless. Second, counter-examples are known to the Ponzano-Regge conjecture, like Bing's house with two rooms considered in \cite{barrett PR}. Third, and more importantly, it does not hold in higher dimensions: when $\Gamma$ is the dual cell-complex to a triangulated $n$-manifold with $n\geq 4$, it is easy to see that $\Omega(\Gamma,G)$ is {\it not} given by the number of $(n-3)$-simplices. (For $n=4$, for instance, a $1$-$5$ Pachner move on the triangulation immediately generates a foam violating this conjecture \cite{ooguri}).


Recently, Gurau \cite{colored gft} has considered a special class of foams, which he calls `colored', and gave a graph-theoretic definition of `bubble' and `bubble homology' for these foams. This is the language in which the results of Ben Geloun et al. in \cite{ben geloun abelien} were framed. Although interesting in itself, this Gurau homology is somewhat idiosyncratic, and it seems preferable to use more stantard, and more general concepts to express the notion of `bubble'.

It is enlightening in this respect to look at the linearization of \eqref{bubble group variables} in the neighbourhood of the trivial connection. This gives
\be\label{linearized}
\sum_{f\in F_v}\, \sum_{e\in\pp f} [e:f]\,X_e = 0, 
\ee
for any choice of $X_e\in\mathfrak{g}$. This observation suggests to consider the cellular homology of $\Gamma$ with coefficients in $\mathfrak{g}$
\be
0\longrightarrow \mathfrak{g}^F\overset{\pp_2(\Gamma,\alg)}{\longrightarrow} \mathfrak{g}^E\overset{0}{\longrightarrow}\mathfrak{g}\longrightarrow 0
\ee
and its dual cohomology
\be
0\longleftarrow \mathfrak{g}^{F}\overset{0}{\longleftarrow}\mathfrak{g}^E\overset{\delta_1(\Gamma,\alg)}{\longleftarrow}\mathfrak{g}\longleftarrow 0
\ee
as the natural setting for powercounting considerations. Here $\pp_1(\Gamma,\alg)$ and $\delta_0(\Gamma,\alg)$ vanish because $\Gamma$ has a single vertex, and $\delta_{1}(\Gamma,\alg)$ and $\pp_{2}(\Gamma,\alg)$ are related to the usual cellular operators with integer coefficients $\delta_{1}(\Gamma,\mathbb{Z})$ and $\pp_{2}(\Gamma,\mathbb{Z})$ by
\begin{align}
\delta_{1}(\Gamma,\alg)&=\delta_{1}(\Gamma,\mathbb{Z})\otimes\id_{\alg}\nonumber\\
\pp_{2}(\Gamma,\alg)&=\pp_{2}(\Gamma,\mathbb{Z})\otimes\id_{\alg}.
\end{align}
In this language, the equation (\ref{linearized}) is simply $\ker\, \pp_2(\Gamma,\mathfrak{g}) \neq \{0\}$, i.e. $H_2(\Gamma,\mathfrak{g}) \neq \{0\}$. Thus, we propose to define a bubble as a $2$-cycle, and the number of bubbles as the second Betti number 
\be
b_2(\Gamma,\alg)=\bigl(\dim G\bigr)\, b_2(\Gamma).
\ee 
Notice that this is definition is purely topologogical, in the sense that the structure group enters only via the multiplicative constant $(\dim G)$. 
Is the divergence degree $\Omega(\Gamma,G)$ given by $b_2(\Gamma,\alg)$? Answering this question is the purpose of this note.

\section{Homological powercounting}\label{divergence}

The ill-defined amplitude \eqref{amplitude} can be conveniently regularized by replacing the delta functions with heat kernels\footnote{Recall that $K_{\tau}$ is the solution of the heat equation on $G$ $$(\pp_{\tau}-\Delta)K_{\tau}=0$$ with initial condition $\u{\tau\rightarrow 0}{\lim}K_{\tau}(g)=\delta(g)$.} $K_{\tau}$ on $G$, and considering the $\tau\rightarrow 0$ limit. Using the standard small-time asymptotics
\be
K_{\tau}(g)\u{\tau\rightarrow 0}{\sim}(4\pi\tau)^{-\f{\dim G}{2}}e^{-\f{\vert g\vert^2}{4\tau}},
\ee
where $\vert g\vert$ is the Riemannian distance between $g$ and the unit $1$ of $G$, and setting
\be
\Lambda_{\tau}=(4\pi\tau)^{-\f{1}{2}}
\ee
as the cutoff, we see that this yields in the $\tau\rightarrow 0$ limit the Laplace-type integral
\be
\mathcal{Z}_{\tau}(\Gamma,G)\u{\tau\rightarrow 0}{\sim}\Lambda_{\tau}^{(\dim G)F}\int_{\mathcal{A}}dA\ e^{-\f{S_{\tau}(A)}{4\tau}},
\ee
with the action
\be
S(A)=\sum_{f\in\Gamma_2}\vert H_f(A)\vert^2.
\ee
It would seem, therefore, that the divergence degree of $\mathcal{Z}_{\tau}(\Gamma,G)$ is determined by the behaviour of the Hessian of the action $S$ along its critical set, which is of course the set of {\it flat} connections $\F$. 
This first intuition is deceiving, however, because in general $\F$ is not a manifold \cite{goldman et karshon}, and the critical points are degenerate. It remains that, when $H$ (and hence $S$) has isolated critical points, perhaps modulo some group action, Laplace's method can still be successful. Indeed, this typically happens in two cases: when $\Gamma$ is simply connected, and when $G$ is Abelian. In both cases, this line of reasoning yields the exact divergence degree -- which, as anticipated, is indeed given by the second Betti number of $\Gamma$:
\be\label{degree}
\Omega(\Gamma,G)=\bigl(\dim G\bigr)\, b_2(\Gamma).
\ee
Moreover, when $\Gamma$ admits a unique flat $G$-connection, the same method still gives lower and upper bounds on $\Omega(\Gamma,G)$.

In these considerations, the following observation will be key:  {\it the differential of the curvature map at the trivial connection is the first cellular coboundary operator of $\Gamma$ with coefficients in $\mathfrak{g}$},
\be \label{key obs}
dH_{\bf 1}=\delta_{1}(\Gamma,\mathfrak{g}).
\ee
In particular, the number of bubbles og $\Gamma$ is given by
\be
b_{2}(\Gamma,\alg)=(\dim G)F-\rk dH_{\bf 1}
\ee
where $\rk$ denotes the rank of a linear map.

\subsection{Simply connected foams}

A simply connected foam admits a unique flat connection, the trivial one: that is, in this case, $S$ has a unique global minimum at $1\in G^E$. The Hessian of $S$ there is the bilinear map on $\mathfrak{g}^E$ defined by
\be
\operatorname{Hess}(S)_{\bf 1}(X,Y)=\la dH_{\bf 1}(X),dH_{\bf 1}(Y)\ra_{\mathfrak{g}^F},
\ee
where $\la\cdot,\cdot\ra_{\mathfrak{g}^F}$ is an invariant inner product in $\alg^F$. By the Hurewicz theorem, we know that $\pi_1(\Gamma)=\{{\bf 1}\}$ implies $H^1(\Gamma)=\ker\delta_1(\Gamma)/\im\delta_0(\Gamma) = 0$, and since $\delta_0(\Gamma,\alg)$ is identically zero (because $\Gamma$ has a single vertex),
\be \label{zero ker delta1}
\ker dH_{\bf 1} = \{0\}.
\ee
Hence, the map $\operatorname{Hess}(S)_{\bf 1}$ is non-degenerate, and the standard Laplace approximation gives
\begin{align}
\int dA\ e^{-\f{S_{\tau}(A)}{4\tau}} &\u{\tau\rightarrow 0}{\sim} \int_{\mathfrak{g}^E} dX\ \exp\bigl(-\f{\Vert dH_{\bf 1}(X)\Vert^2_{\mathfrak{g}^F}}{4\tau}\bigr), \nonumber\\
&\u{\tau\rightarrow 0}{\sim} \Lambda_\tau^{-\rk dH_{\bf 1}}\ \bigl(\det \operatorname{Hess}(S)_{\bf 1}\bigr)^{-1/2}.
\end{align}
Here the Hessian determinant is evaluated in an orthonormal basis of $\mathfrak{g}^E$. Combining the exponent of $\Lambda_{\tau}$ coming from the asymptotics of $K_{\tau}$ with the Gaussian contribution $\rk dH_{\bf 1}$, and using the cohomological identity

\be
\rk dH_{\bf 1}=\rk\delta_{1}(\Gamma,\alg)=(\dim G)(F-b_{2}(\Gamma)),
\ee
we obtain

\be
\mathcal{Z}_{\tau}(\Gamma,G) \u{\tau\rightarrow 0}{\sim} \Lambda_{\tau}^{(\dim G) b_2(\Gamma)}\ \bigl(\det \operatorname{Hess}(S)_{\bf 1}\bigr)^{-1/2}.
\ee
In fact, this result is slightly more general. Indeed, it relies on two ingredients only: flat connections where the condition \eqref{key obs} holds, and the non-degeneracy condition \eqref{zero ker delta1}.\footnote{The requirement that flat connections be isolated is actually enforced by the non-degeneracy condition on the Hessian, by the Morse lemma.} The first condition is ensured if flat connections live in the center of $G^E$, since then
\be
dH_{A}=\delta_{1}(\Gamma,\alg)\circ\theta_{A},
\ee
with $\theta$ the Maurer-Cartan form on $G^E$, and therefore $\rk dH_{A}=\rk\delta_{1}(\Gamma,\alg)$. For instance, if $G=\SU$, the two conditions are satisfied if $\pi_1(\Gamma)$ is a dihedral group $D_{2n}$, a symmetric group $S_n$, the alternating group $A_5$, or to $GL(2,\mathbbm{Z})$.Using some standard presentations of these groups\footnote{Precisely, they admit the following presentations: $D_{2n}=\langle r,f \vert r^n=f^2=(rf)^2={\bf 1}\rangle, S_n=\langle (\sigma_i)_{i=1,\dotsc,n-1}\vert \sigma_i={\bf1}, \sigma_i\sigma_j = \sigma_j \sigma_i\ \text{if}\ j\neq i\pm 1, \sigma_i\sigma_{i+1}\sigma_i = \sigma_{i+1}\sigma_i\sigma_{i+1}\rangle, A_5 = \langle s, t \vert s^2 = t^3 = (st)^5 ={\bf 1}\rangle$ and $GL(2, \mathbbm{Z}) = \langle j, a , b \vert j^2 = (ja)^2 =(jb)^2 = (aba)^4 ={\bf1}, aba = bab\rangle$. The fact that the representations of these groups into $\SU$ are very simple can be traced to the fact that the equation $g^2={\bf 1}$ is solved in $\SU$ only by $g=\pm {\bf 1}$.}, it can be checked, through direct computations, that flat connections live in the center $\{-{\bf 1},{\bf 1}\}^E$. Furthermore, non-degeneracy of the Hessians follows from looking at the abelianization of $\pi_1(\Gamma)$, from which it can be concluded that the spaces $H^1(\Gamma)$ are trivial.

\subsection{Abelian structure group}

When $G$ is Abelian, another simplification arises: the curvature map $H:G^E\rightarrow G^F$ is a Lie group homomorphism. This means that $H$ can be factored through its kernel -- the flat connections -- and thus be made injective. In other words, in the Abelian case, the gauge fixing procedure (reduction of a maximal tree in $\Gamma$) can be superseded by the global factorization of the flat connections. Here is how this goes. 

Let $\bar{H}$ denote the quotient of $H$ trough its kernel, and $d\bar{A}$ the Haar measure on (the compact Lie group) $Q=G^{E}/\ker H$. Then we have 
\be
S(A)=\sum_{f\in\Gamma_2}\vert\bar{H}_f(\bar{A})\vert^2,
\ee
where $\bar{A}$ is the equivalence class of $A$ modulo $\ker H$. This defines a function $\bar{S}$ on $G^{E}/\ker H$ having a {\it unique} global minimum at ${\bf 1}\in Q$. Up to exponentially suppressed corrections, the Laplace integral can therefore be restricted to a neighbourhood $U$ of ${\bf 1}$ where
\be
\bar{S}(\bar{A})=\Vert d\bar{H}_{\bf 1}(\log\bar{A})\Vert_{\mathfrak{g}^{F}}^2.
\ee
To deal properly with the degeneracy of the Hessian, we introduce Riemannian normal coordinates on $U$ adapted to the decomposition
\be
\mathfrak{q}=\ker d\bar{H}_{\bf 1}\oplus(\ker d\bar{H}_{\bf 1})^{\perp},
\ee
and integrate in the direction of $\ker d\bar{H}_{\bf 1}$. This gives
a constant $C$. Next, we perform the Gaussian integration in the orthogonal direction, observing that $\rk\bar{H}=\rk H$, and from $dH_{\bf 1}=\delta_{1}(\Gamma,\mathfrak{g})$ we conclude that 
\be
\mathcal{Z}_{\tau}(\Gamma,G)\u{\tau\rightarrow 0}{\sim}C\Lambda_{\tau}^{\Omega(\Gamma,G)}\det(d\bar{H}_{\bf 1})^{-1/2}_{\vert(\ker d\bar{H}_{\bf 1})^{\perp}}
\ee
with $\Omega(\Gamma,G)$ as in (\ref{degree}). That is, in the Abelian case, the divergence degree is given by the number of bubbles of $\Gamma$ {\it whatever its topology}.

\subsection{Single degenerate flat connection}

Let us eventually point out some difficulties arising when the critical points of the action are degenerate. (For simplicity, we assume that $\Gamma$ admits a single flat connection up to gauge transformations.) As before, the tangent space at the identity has a natural splitting $\mathfrak{g}^E = \ker dH_{\bf 1} \oplus (\ker dH_{\bf 1})^\perp$. However, unlike the Abelian case, $S$ will not just be quadratic on $\ker dH_{\bf 1}$, and one has to further expand it along these directions. Writing $S\approx \operatorname{Hess}(S)_{{\bf 1}|(\ker dH_{\bf 1})^\perp} + T_{\ker dH_{\bf 1}}$:
\begin{align}
\mathcal{Z}_{\tau}(\Gamma,G) \u{\tau\rightarrow 0}{\sim} \Lambda_{\tau}^{(\dim G)F} &\int_{(\ker dH_{\bf 1})^\perp} dX\ e^{-\f{\Vert dH_{\bf 1}(X)\Vert^2_{(\ker dH_{\bf 1})^\perp}}{4\tau}} \nonumber\\
&\times \int_{\ker dH_{\bf 1}} dY\ e^{-\f{T(Y)}{4\tau}},
\end{align}
where $T(Y)$ is at least quartic in $Y$.

Despite the possibly complicated behaviour of $T$, one can easily get some bounds on the degree of divergence $\Omega(\Gamma,G)$ in that situation. Indeed, the basic idea is that the integral over the kernel of $dH_{\bf 1}$, after some truncation of the expansion of $T$, reduces the degree of divergence by providing negative powers of $\Lambda_\tau$. This can be seen for instance when $T$ is a monomial of total degree $k$. Then, rescaling $Y$ by $\tau^{-1/k}$ produces a factor $\Lambda_\tau^{-2\f{\dim \ker dH_{\bf 1}}{k}}$. The dimension of $\ker dH_{\bf 1}$ being fixed, this means that the most divergent case corresponds to the vanishing of $T$ at all orders. Since $G$ is compact, the integral over $Y$ does not bring about additional divergences. Thus, the divergence degree is always smaller than the one obtained for $T=0$:
\begin{align}
\Omega(\Gamma,G) &\leq \bigl(\dim G\bigr) F- \rk dH_{\bf 1},\nonumber\\
&\leq \bigl(\dim G\bigr) b_2(\Gamma).
\end{align}
Furthermore, the least divergent situation is for $T$ being exactly quartic on $\ker dH_{\bf 1}$. Then, a simple rescaling of $Y$ by $\tau^{-1/4}$ leads to the following bound on the divergence degree:
\be
\Omega(\Gamma,G) \geq \dim G\ \bigl(b_2(\Gamma) - \f 12 \dim\ker \delta_1(\Gamma)\bigr).
\ee
Since $\delta_0(\Gamma)$ is identically zero, the dimension of $\ker\delta_1(\Gamma)$ is the Betti number $b_1(\Gamma)$. Introducing the Euler characteristic of $\Gamma$, $\chi(\Gamma) = b_2-b_1+1$, we finally get the following bounds:
\be
\f 12\bigl(b_2(\Gamma) + \chi(\Gamma) -1\bigr) \leq \f{\Omega(\Gamma,G)}{\dim G} \leq b_2(\Gamma).
\ee

Although the above bounds only depend on $\Gamma$ and do not exhibit any entanglement between $\Gamma$ and $G$, this analysis already shows that the exact degree $\Omega(\Gamma,G)$ will depend on $G$ in a more sophisticated way than the simple scaling by $(\dim G)$. By the BCH formula, one sees that $T_{\ker dH_{\bf 1}}$ depends on the (non-)commutative structure of the Lie algebra $\mathfrak{g}$. In particular $T$ identically vanishes as soon as $G$ is Abelian, consistently with the result of the previous section.

\section{Conclusion}

We have shown in this note that, in some special cases, the intuition that the divergence degree of a foam depends on its topology only (up to the multiplicative constant $\dim G$) is correct. But, in our opinion, more interesting is the realization that {\it this is not true in general}. Technically, we saw that this is because the flat configurations form an extended set, along which the action $S(A)$ has a degenerate Hessian. An example of this phenomenon is provided by the weak-coupling limit of $2$d Yang-Mills theory: for genus $g\geq 2$, the partition function is well-known to converge as $\tau\rightarrow 0$ for $G=\SU$, and to diverge for Abelian groups.

Still, it is possible to adapt the cohomological analysis to this case, by making the cochain complex {\it local}. For every flat connection $\phi\in\F$, $dH_{\phi}$ defines a {\it twisted} coboundary operator \cite{goldman et karshon}. This local cochain complex is well-known in two dimensional Yang-Mills theory \cite{witten, forman}, and similar analyses have been performed for BF theories in the continuum \cite{blau thompson BF, horowitz, gegenberg}. In our framework, this line of reasoning yields a {\it local divergence degree} $\omega(\Gamma,G;\phi)$, such that $0<\u{\Lambda\rightarrow 0}{\lim} \vert\Lambda^{-\omega(\Gamma,G;\phi)}\mathcal{Z}_{\tau}(\Gamma,G;\phi)\vert<\infty$ where
\be
\mathcal{Z}_{\tau}(\Gamma,G,\phi)=\int_{U_{\phi}}dA\prod_{f\in\Gamma_2}K_{\tau}(H_f(A)),
\ee
for some neighbourhood $U_{\phi}\subset\mathcal{A}$ of $\phi$, given (or bounded) by the second Betti number in this twisted cohomology \cite{twisted}. To infer the global divergence degree $\Omega(\Gamma,G)$ from this local analysis, one should then study the behaviour of this twisted cohomology as $\phi$ approaches the singularities of $\F$.

In short, the divergence degree for a generic foam $\Gamma$ involves both the topology of $\Gamma$ and the non-Abelian structure of $G$ in a rather subtle way -- indeed much subtler than the notion of `number of bubbles' would suggest {\it a priori}.

\section*{Acknowledgements}

We are indepted to Jacques Magnen and Thomas Krajewski for the suggestion to study the Abelian case first, and to Razvan Gurau, Joseph Ben Geloun, Vincent Rivasseau and Henrique Gomes for useful discussions on homology and powercounting. We also thank collectively the Marseille quantum gravity group for their support.


\end{document}